\documentclass[epsf,twocolumn,preprintnumbers]{revtex4}
\usepackage{epsfig,graphics}
\usepackage{dcolumn}
\usepackage{bm}
\usepackage{amsmath}
\usepackage{xcolor}
\usepackage{comment}

\definecolor{dkred}{rgb}{0.6,0.0,0.0}
\definecolor{dkblue}{rgb}{0.123,0.247,0.554}

\def\Re{\mathop{\rm Re}}
\def\Im{\mathop{\rm Im}}

\begin{document}
\global\long\def\bra#1{\left\langle #1\right|}%
\global\long\def\ket#1{\left|#1\right\rangle }%
\global\long\def\braket#1#2{\langle#1|#2\rangle}%
\title{Spin-orbit coupling and Kondo resonance in Co adatom on Cu(100) surface: DFT+ED study}

\author{A.  B.  Shick}
\affiliation{Institute of Physics, Czech Academy of Science,  Na Slovance 2, CZ-18221 Prague, Czech Republic}
\email{shick@fzu.cz}

\author{M. Tchaplianka}
\affiliation{Institute of Physics, Czech Academy of Science,  Na Slovance 2, CZ-18221 Prague, Czech Republic}

\author{A. I. Lichtenstein}
\affiliation{Institute  of  Theoretical  Physics,  University  of  Hamburg,  20355  Hamburg,  Germany}
\affiliation{European  X-Ray  Free-Electron  Laser  Facility,  Holzkoppel  4,  22869  Schenefeld,  Germany}

\date{\today}

\begin{abstract}
We report density functional theory plus exact diagonalization of the multi-orbital Anderson impurity model calculations  for the Co adatom on the top of Cu(001) surface. 
For the Co atom $d$-shell occupation $n_d \approx$ 8, a singlet many-body ground state and Kondo resonance are found, when the spin-orbit coupling
is included in the calculations.  The differential conductance is evaluated in a good agreement with the scanning
tuneling  microscopy measurements.  The results illustrate the essential role which the spin-orbit coupling
is playing in a formation of Kondo singlet for the multi-orbital impurity in low dimensions.
\end{abstract}

\maketitle

\section{Introduction}
The electronic nanometer scaled  devices require the atomistic control of their behaviour
governed by the electron correlation effects. One of the most famous correlation phenomena
is  the Kondo effect originating from screening of the local magnetic moment by the Fermi sea
of conduction electrons, and resulting in a formation of a singlet ground state~\cite{Mahan}.
Historically the Kondo screening was detected as a resistance increase below a characteristic
Kondo temperature $T_K$ in dilute magnetic alloys~\cite{monod1967}. Recent advances in scanning
tuneling  microscopy (STM) allowed observation of the Kondo phenomenon on the atomic scale,
for atoms and molecules at surfaces~\cite{Knorr2002,Zhao2005}. 
In these experiments, an enhanced conductance near the Fermi level ($E_F$) is found due to the formation
of  a sharp Abrikosov-Suhl-Kondo~\cite{abrikosov,suhl,nagaoka} resonance in the electronic 
density of states (DOS).

One case of the Kondo effect the most studied experimentally and theoretically is that of a Co adatom on the metallic 
Cu substrate~\cite{Knorr2002,Wahl2004,Surer2012,Valli2020}. The experimental STM spectra display sharp peaks 
at zero bias, or so called "zero-bias" anomalies, similar to the Fano-resonance~\cite{Fano1961} 
found in the atomic physics, which are associated with the Kondo resonance. 
The theoretical description of the Kondo screening in multiorbital $d$ manifold is difficult since the whole $d$ shell is likely
to play a role. Very recently, theoretical electronic structure of the Co atom on the top of Cu(100) was considered~\cite{Valli2020} using numerically exact
continuous-time quantum Monte-Carlo (CTQMC) method~\cite{Rubtsov2005} to solve the multiorbital single impurity Anderson model~\cite{Hewson}
(SIAM) together with the density-functional theory~\cite{DFT} as implemented in the W2DYNAMICS 
package~\cite{Parragh2012,Wallerberger2019}. However, the spin-orbit coupling (SOC) was neglected.
The peak in the DOS at $E_F$ was obtained in these calculations, and was interpreted  as a signature of the Kondo resonance.  

Alternative interpretation  was proposed~\cite{Lounis2020} which is based on the spin-polarized time-dependent DFT in 
conjunction with many-body perturbation theory.
These authors claim that the "zero-bias" anomalies are not necessarily related to the Kondo resonance,
and are connected to interplay between the inelastic spin excitations and the magnetic anisotropy.
Thus the controversy exists concerning the details of the physical processes underlying the Kondo screening
in Co@Cu(100). In this work, we revisit Co@Cu(100) case making use of the combination of DFT with the exact 
diagonalization of multiorbital SIAM (DFT+ED) including SOC. We demonstrate that SOC plays crucial role in 
formation of the singlet ground state (GS) and the Kondo resonance. 

\section{Methodology: DFT + Exact Diagonalization}
\begin{figure*}[!htbp]
\centerline{\includegraphics[angle=0,width=1.0\columnwidth,clip]{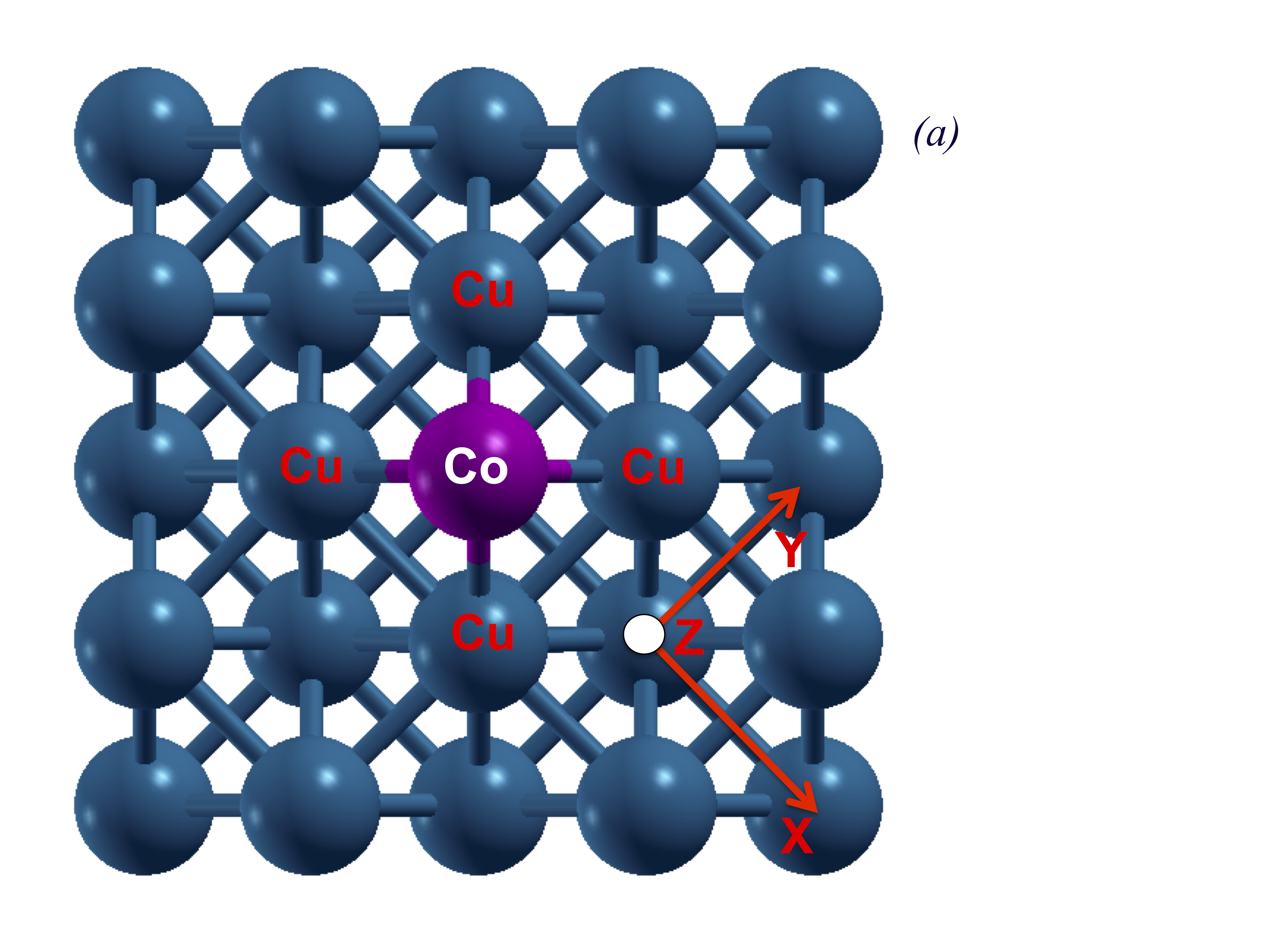}}
\centerline{\includegraphics[angle=0,width=1.0\columnwidth,clip]{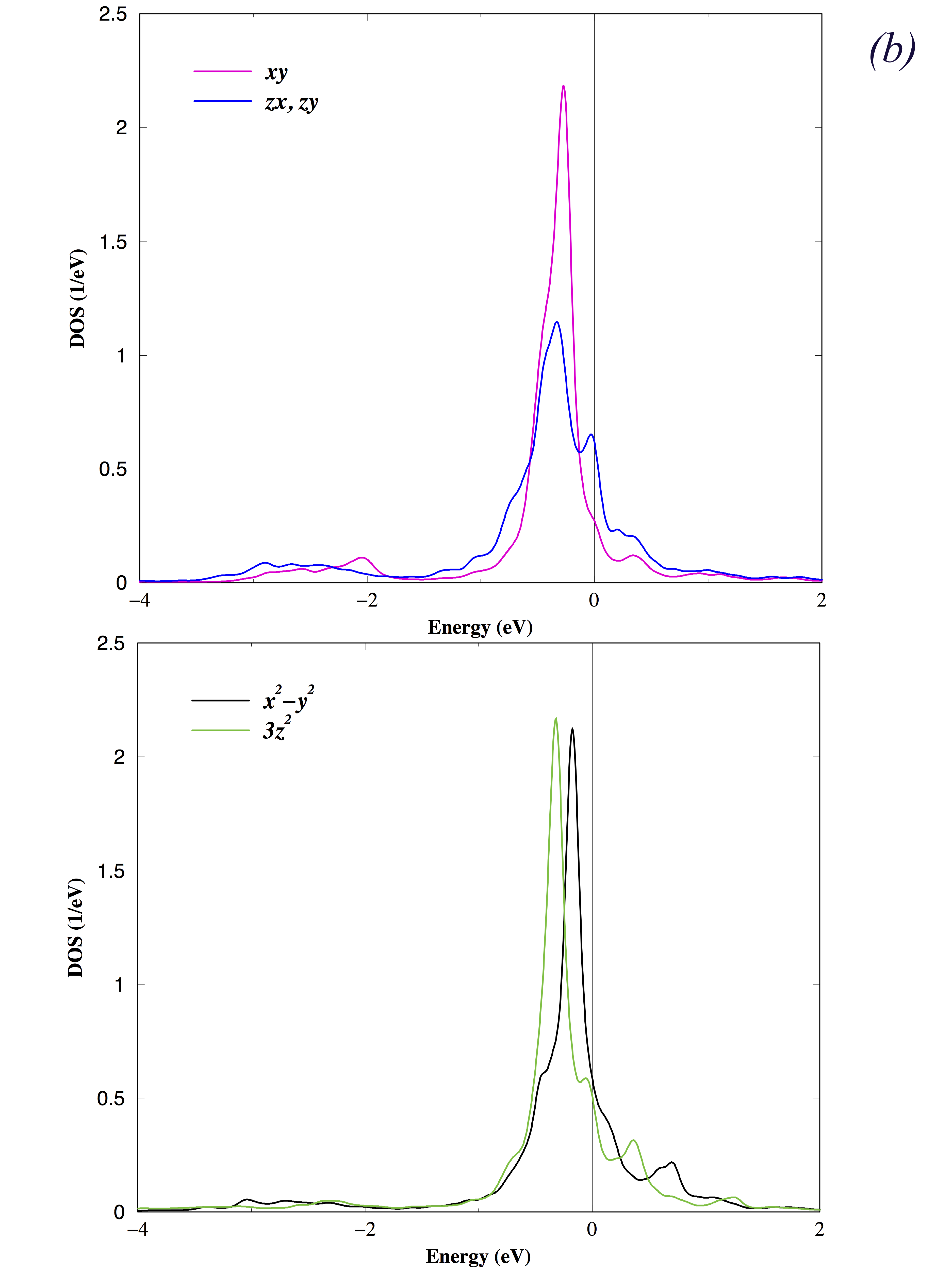}
\includegraphics[angle=0,width=1.0\columnwidth,clip]{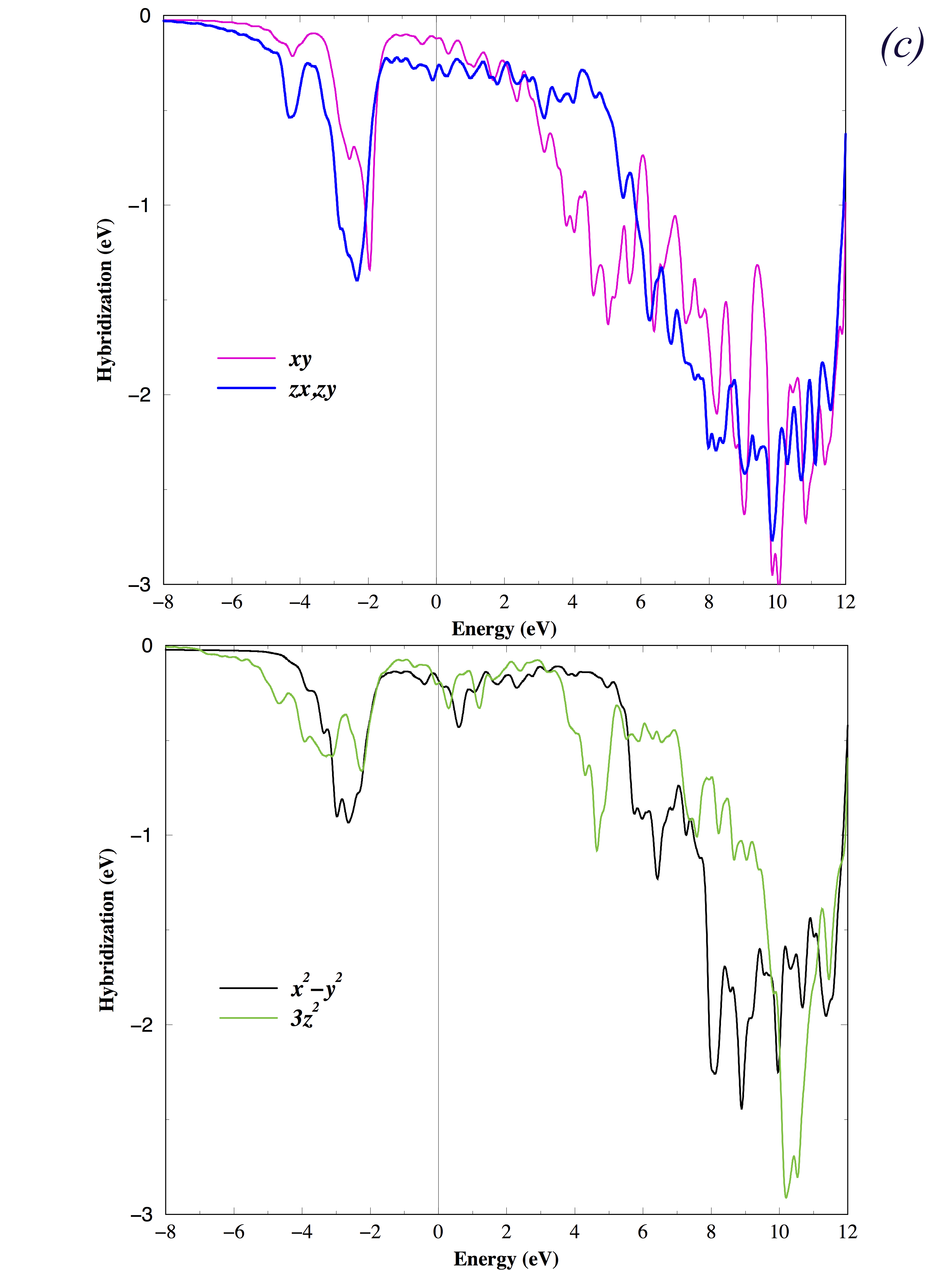}}
\caption{The ball model (top view) for Co@4Cu$_8$] supercell. The specific choice of the Cartesian reference frame
is show. With this choice, the local Green's function without SOC becomes diagonal in the basis of cubic harmonics 
$m = \{xz, yz, xy, x^{2}-y^{2}, 3z^{2}-r^{2}\}$ (a);
orbitally resolved DOS (b);  orbitally resolved hybridization $\Im \Delta$ (c)
for the Co adatom on Cu(001),
}
 \label{fig1}
\end{figure*}

The exact diagonalization (ED) method is based on a numerical solution of the multi-orbital Anderson impurity
model (AIM)~\cite{Hewson}.  The continuum of the bath states
is discretized.  The five $d$-orbitals AIM with the full spherically symmetric Coulomb interaction,
a crystal field (CF), and SOC is written as,
\begin{align}
\label{eq:hamilt}
H =  & \sum_{km\sigma}
   \epsilon_{km} b^{\dagger}_{km\sigma} b_{km\sigma} 
 + \sum_{m\sigma}\epsilon_d d^{\dagger}_{m \sigma}d_{m \sigma}
\nonumber \\ 
& + \sum_{mm'\sigma\sigma'} \bigl(\xi {\bf l}\cdot{\bf s} 
  +\mbox{\boldmath $\Delta_{\rm CF}$}
 \bigr)_{m m'}^{\sigma \; \; \sigma'}
  d_{m \sigma}^{\dagger}d_{m' \sigma'} \nonumber \\  
& +  \sum_{km\sigma} \Bigl( V_{km}
  d^{\dagger}_{m \sigma} b_{km \sigma} + {h.c.}
  \Bigr) \\
&   + \frac{1}{2}\sum_{ {m m' m''  m''' \sigma \sigma'}} 
  U_{m m' m'' m'''} d^{\dagger}_{m\sigma} d^{\dagger}_{m' \sigma'}
  d_{m'''\sigma'} d_{m'' \sigma}.
\nonumber  
\end{align}

The impurity-level position $\epsilon_d$ which yield the desired $\langle n_d \rangle$, and the bath
energies $\epsilon_{km}$ are measured from the chemical potential $\mu$, that was set to zero. 
The SOC $\xi$ parameter  specifies the strength of the spin-orbit coupling, whereas {\boldmath $\Delta_{\rm CF}$} matrix describes CF acting on the impurity. 
The hybridization $V_{mk}$ parameters describe the coupling of substrate to the impurity orbitals. These parameters are determined from DFT calculations,
and their particular choice will be described below.

The last term in Eq.(\ref{eq:hamilt}) represents the Coulomb interaction. The Slater integrals $F_0$= 4.00 eV, $F_2$=7.75 eV, and  $F_4$= 4.85 eV  are used for the Coulomb 
interaction~\cite{Surer2012, Valli2020}. They correspond to the values for the Coulomb $U$ = 4 eV and exchange
$J$ = 0.9 eV for Co which are in the ballpark
of commonly accepted $U$ and $J$ for transitional 3$d$-metals.

The DFT calculation  were performed on a supercell of four Cu(100) layers, and the Co adatom   followed by four empty Cu layers modeling the vacuum. Fig.~\ref{fig1}A shows ball model of the Co@[4Cu$_8$] supercell employed for the adsorbate atop of Cu . 
The structure relaxation was performed employing the VASP method~\protect\cite{kre1996} together with the  generalized gradient
approximation (GGA) to spin-polarized DFT without SOC. 
The adatom-substrate distance as well as the atomic positions within two Cu(100) layers underneath were allowed to relax.
The relaxed distance between the Co adatom in a fourfold hollow position and the first Cu substrate layer of 2.91 $a.u.$ is in a good agreement
with previously reported value of 2.87 $a.u$~\cite{Valli2020}. 

In order to obtain the bath parameters in the AIM Hamiltonian Eq.(~\ref{eq:hamilt}) we make use of the
recipes of the dynamical mean-field theory (DMFT)~\cite{KG,LK}, and employ the DFT(LDA)  local Green's function $G_0(z)$
\begin{equation}
[G_0(z)]_{\gamma_1 \gamma_2} = \frac{1}{V_{\rm BZ}}
\int_{\rm BZ}{\rm d}^3 k \,\bigl[z+E_F-H_{\rm DFT}({\bf
k})\bigr]^{-1}_{\gamma_1 \gamma_2}\,,
 \label{eq:gf}
\end{equation}
calculated with help of the full-potential linearized augmented plane wave method 
(FLAPW)~\cite{Wimmer1981,Mannstadt1997}, in order to define the parameters for the Eq.(\ref{eq:hamilt}) .  
Here, the energy $z$ is counted from the
Fermi energy $E_F$, and the index $\gamma \equiv l m \sigma$ marks the $d$-orbitals in the MT-sphere of the Co adatom.
Note that the non-spin-polarised LDA is used to extract the hybridization function $\Delta(z)$. The orbitally resolved density of states (DOS) together with the
hybridization function $\Im \Delta$ are shown in Fig.~\ref{fig1}B,C. They are compatible with the results of Ref.~\cite{Jacob2015}.
Further details of constructing the discrete bath model are given in Appendix A.
The fitted bath parameters are shown in Table~\ref{tab:1}. These parameters are used to build the AIM Hamiltonian Eq.(~\ref{eq:hamilt}). 

The SOC parameter $\xi$= 0.079 eV is taken from LDA  calculations in a standard way,

$$\xi = \int_{0}^{R_{MT}} dr r \frac{1}{2(Mc)^2} \frac{dV(r)}{dr} (u_{l}(r))^2  \; ,$$

making use of the radial solutions $u_l$ of the Kohn-Sham-Dirac scalar-relativistic equations~\cite{MPK1980}, the
relativistic mass $M = m + (E_l - V(r)) / {2 c^2}$ at an appropriate energy $E_l$, and the radial derivative of spherically-symmetric
part of the LDA potential. 

\section{Results and Discussion}
The total number of electrons $N$, and the $d$-shell occupation are controlled 
by the $\epsilon_d$ parameter. It has a meaning of the chemical potential $\mu = -\epsilon_d$ in Eq.~(\ref{eq:hamilt}).
In DMFT it is quite common to use $\mu = V_{dc}$, the spherically-symmetric double-counting
which has a meaning of the mean-field Coulomb energy of the $d$-shell, and to use standard 
(AMF)  $V_{dc} =  (U/2 \; {n_d } + \frac{2l}{2(2l+1)} \; (U-J) \; {n_d})$~\cite{AZA1991} form, or the fully localized limit (FLL)
$V_{dc} =  (U-J)/2 \;  ({n_d} - 1)$~\cite{solovyev1994}.  Since precise definition of $n_d$ depends on the choice of the
localized basis, we adopt a strategy of Ref.~\cite{Surer2012}, and consider a value of $\mu$ as a parameter.

\subsection{Co in the bulk Cu}

At first, we consider the Co impurity in the bulk Cu making use of  the CoCu$_{15}$ supercell model.
DFT+ED calculations for different values of $\mu$ in a comparison with previous DFT+CTQMC
relsults~\cite{Surer2012} are  described in details in Ref.~\cite{Tchaplianka2022}. Here, we adjust the 
value of $\mu$   in order to have the Co atom $d$-shell occupation $n_d \approx$ 8. This valence of 
Co in the bulk Co follows from DFT calculations~\cite{Surer2012,Tchaplianka2022}.

Without SOC we found that the value of $\mu=27.4$  corresponds to the $n_d \approx$ 8 occupation.
The GS solution without SOC (see Table~\ref{tab:4}) is the $\ket{\Omega}_{N=30}$ singlet,
and the exited triplet is $\approx$ 0.4 eV higher in the energy. Note that each eigenstate  $\ket{\Omega}_{N}$  
of  Eq.(~\ref{eq:hamilt}) corresponds to an integer $N$ occupation ($d$-shell + bath) since $\hat{N}$ commutes with
Hamiltonian Eq.(~\ref{eq:hamilt}). For each $\ket{\Omega}$,  the  probabilities to find the atomic eigenstates $\ket{n}$ with integer occupation $d^{n}$,
$P_n =  \braket{n}{\Omega} \braket{\Omega}{n}$, and  the $d$-shell occupation $n_d = \sum_n P_n n_{d^{n}}$.

\begin{figure}[!htbp]
\centerline{\includegraphics[angle=0,width=1.0\columnwidth,clip]{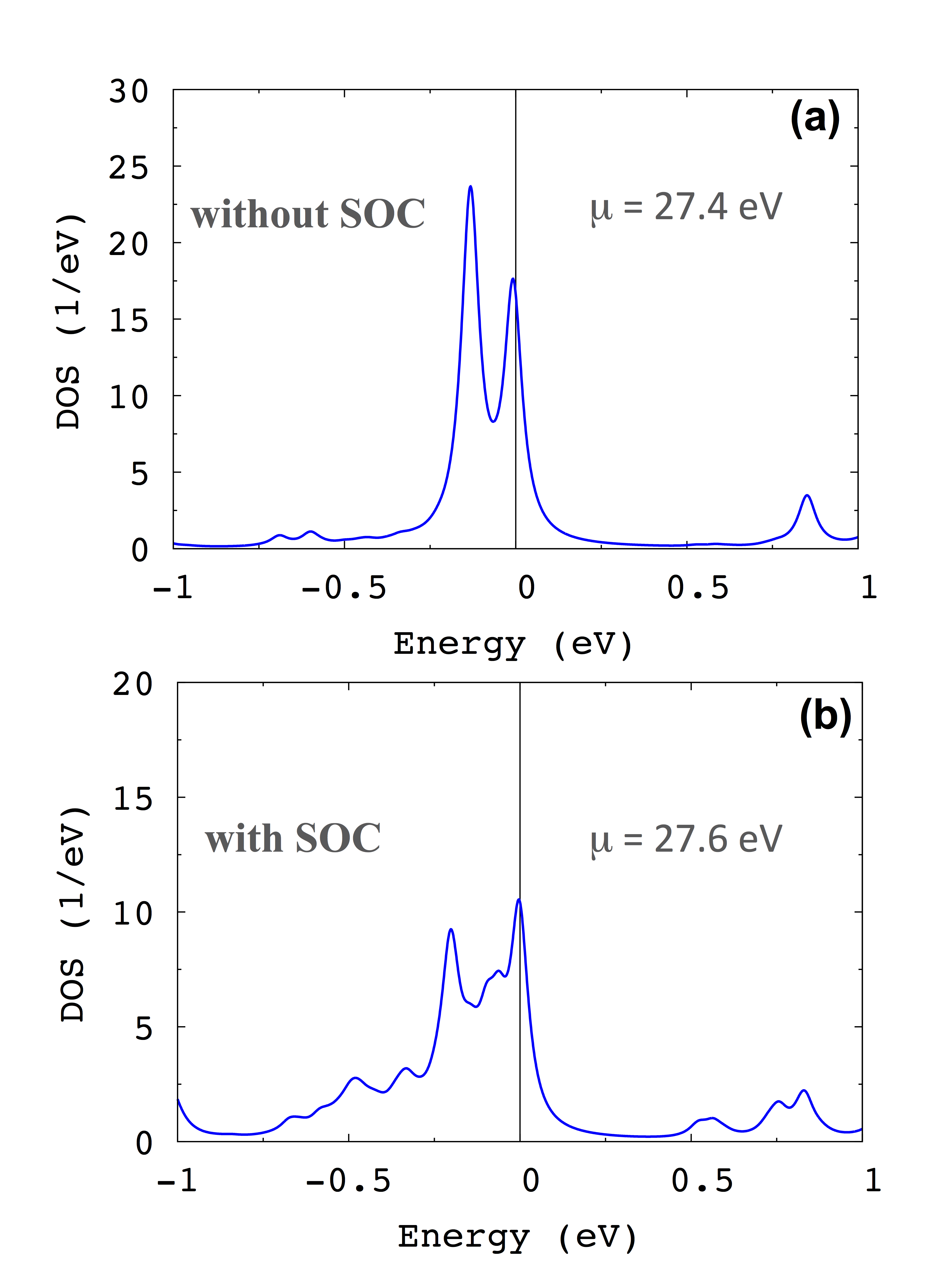}}

\caption{DOS for the Co in the bulk Cu without SOC for  $\mu$= 27.4 eV  (a), 
and with SOC for $\mu$= 27.5 eV  (b).}
 \label{fig:5}
\end{figure}

The corresponding density of $d$-states (DOS)
~\cite{Mahan}:
\begin{eqnarray}
A(\epsilon)=-\frac{1}{\pi Z}{\Im \sum_{\gamma,\alpha,\beta}
  \frac{ \langle \Omega_{\alpha}|c_{\gamma}|\Omega_{\beta}\rangle 
         \langle \Omega_{\beta}|c^{\dag}_{\gamma}|\Omega_{\alpha}\rangle } 
       {\epsilon + i\delta +E_{\alpha}-E_{\beta}} 
    [e^{-\beta E_{\beta}}+e^{-\beta E_{\alpha}}}]  
\label{eq:sdos}
\end{eqnarray}
where the $\alpha,\beta$ run over the eigenstates of Hamiltonian Eq.(~\ref{eq:hamilt}), $\gamma \equiv \{m , \sigma\}$ marks the single particle spin-orbital,
is shown in Fig.~\ref{fig:5}a,  with the peak in DOS very near $E_F$. 

The expectation values of the total $\bra{\Omega}J_z\ket{\Omega}$, orbital $\bra{\Omega}L_z\ket{\Omega}$, and spin $\bra{\Omega}S_z\ket{\Omega}$ angular momenta
for the $\ket{\Omega}_{N=30}$ singlet GS and the exited triplet are shown in Table~\ref{tab:4}. They correspond to a solution of the Kondo model
for localized $S= {1 \over 2}$ anti-ferromagnetically coupled to a single band of conduction electrons~\cite{Yosida1966}. Together with the Kondo peak in DOS  (cf. Fig.~\ref{fig:5}a)
our DFT+ED solution corresponds to the Kondo singlet state.

When SOC is included, and the spin is not a good quantum number, there are a minor changes in the character for $\mu=27.5$ ($n_d \approx 8$), the GS solution $\ket{\Omega}_{N=30}$:
GS is a singlet, and the exited triplet consists of an effective   $\ket{J=1,J_z=-1,0,1}$ degenerate states which are $\approx$ 0.5 eV higher in the
energy. The DOS has a peak in DOS very near $E_F$.  It is seen that  weak 3$d$-shell SOC plays no essential role for the Co impurity in the Cu host.
These calculations show that our DFT+ED approach is capable to reproduce the Kondo singlet for Co in the bulk Cu for $n_d = 8$, in agreement with conclusions of
DFT+CTQMC~\cite{Surer2012}. Also, in agreement with commonly accepted
point of view~\cite{Bergmann1986}, we show that the presence of SOC does not lead to essential modification of a Kondo model.


\begin{table}[!htbp]
\caption{The total number of particles ($d$-shell + bath) $N$, the expectation values $\bra{\Omega}J_z\ket{\Omega}$, $\bra{\Omega}L_z\ket{\Omega}$,$\bra{\Omega}S_z\ket{\Omega}$ angular momenta,
non-zero probabilities $P_{d^n}$ to find the atomic eigenstates $\ket{n}$ with integer occupation $d^{n}$
 for GS and low-energy  excitation energies for different values of $\mu$.}
\label{tab:4}
\centering
\begin{tabular}{cccccccc}
\multicolumn{8}{c}{without SOC}\\
\hline
& Energy (eV) & $J_z$ & $L_z$ & $S_z$ & $P_{d^7}$ & $P_{d^8}$ &$P_{d^9}$\\
\hline
\multicolumn{8}{c}{$\mu$=27.4 eV, $n_d$= 8.05}\\
\hline
$N$=30 & -148.5822 & 0. & 0. & 0.    & 0.20      &0.51       & 0.26    \\
             & -148.1014 & 0.53 & 0 & 0.53    & 0.22      &0.55       & 0.20    \\
             & -148.1014 & 0 & 0 & 0    & 0.22      &0.55       & 0.20    \\
             & -148.1014 & -0.53 & 0 & -0.53    & 0.22      &0.55       & 0.20    \\
\hline
\multicolumn{8}{c}{with SOC}\\
\hline
& Energy (eV) & $J_z$ & $L_z$ & $S_z$& $P_{d^7}$ & $P_{d^8}$ &$P_{d^9}$\\
\hline
\multicolumn{8}{c}{$\mu$=27.5 eV, $n_d$= 7.99}\\
$N$=30 & -149.4028 & 0. & 0. & 0.    & 0.19      &0.51       & 0.27    \\
             & -148.9296 & 0.94 & 0.49 & 0.45    & 0.21      &0.55       & 0.21    \\
             & -148.9296 & 0. & 0. & 0.    & 0.21      &0.55       & 0.21    \\
             & -148.9296 & -0.94 & -0.49 & -0.45    & 0.21      &0.55       & 0.21    \\ 
\hline
 \end{tabular}
\end{table}

\subsection{Co on Cu(001)}

Now we turn to a salient aspect of our investigation, the Co adatom on Cu(001) surface.
Considering a value of $\mu$ as a parameter, we analyse the ground state (GS) of  Eq.(~\ref{eq:hamilt}) with and without SOC for different values of  $\mu$.
Making use of  grand-canonical averages at low temperature $k_{\rm B}T= \beta^{-1}=(1/500)$ eV (20K) we calculate 
the expectation values of total number of electrons ($d$-shell + bath)  $\langle N \rangle$,  the charge fluctuation ${(\langle N^2 \rangle - {\langle N \rangle}^2)}^{1 \over 2}$
near the GS,
 the expectation values  of spin ($S$), orbital ($L$) and
total spin-orbital ($J$) moments, and show them in Table \ref{tab:2} together with   
the $d$-shell occupation $n_d$ for the GS, and corresponding 
$P_n$ probabilities, with and without SOC.

\begin{table*}[!htbp]
\caption{The chemical potential $\mu$ (eV), 
the occupation $\langle N \rangle$, fluctuation ${(\langle N^2 \rangle - {\langle N \rangle}^2)}^{1 \over 2}$,
 $n_d$ occupation, 
non-zero probabilities $P_{d^n}$ to find the atomic eigenstates $\ket{n}$ with integer occupation $d^{n}$,  spin, orbital and total moments of the impurity
$d$-shell for different values of $\mu$. Grand-canonical averages are at low temperature $k_{\rm B}T= \beta^{-1}=(1/500)$ eV.}
\label{tab:2}
\centering
\begin{tabular}{ccccccccccccc}
\multicolumn{2}{c}{\bf without SOC}\\
\hline
$\mu$ (eV)& $\langle N \rangle$ &${(\langle N^2 \rangle - {\langle N \rangle}^2)}^{1 \over 2}$& $n_{d}$ & $P_{d^6}$ & $P_{d^7}$ & $P_{d^8}$ &$P_{d^9}$ & $S$ & $L$ & $J$\\
\hline
26              &26.00& 0.00& 7.57 & 0.05 & 0.34 & 0.56& 0.03 & 1.10 & 3.07 & 3.40\\
\hline
27             &26.00&0.01 &7.74 & 0.03 &  0.27 &  0.62 &  0.08 & 1.03 & 3.01 & 3.32\\
\hline
27.4          &26.55&0.50 &7.93 & 0.02 &  0.21 &  0.58 &  0.18  & 0.94 & 2.87 & 3.15\\
 \hline
28            &27.00&0.00 & 8.17 & 0.01 &  0.14 &  0.51 &  0.33 & 0.82 & 2.68 & 2.91\\
\hline
\end{tabular}

\begin{tabular}{cccccccccccc}
\multicolumn{2}{c}{\bf with SOC}\\
\hline
$\mu$ (eV)& $\langle N \rangle$ &${(\langle N^2 \rangle - {\langle N \rangle}^2)}^{1 \over 2}$&$n_{d}$ & $P_{d^6}$ & $P_{d^7}$ & $P_{d^8}$ &$P_{d^9}$ & $S$ & $L$ & $J$ \\
\hline
26              &26.00&0.00& 7.58 & 0.05 & 0.34 & 0.57& 0.04 & 1.09 & 3.07 & 3.89\\
\hline
27             &26.00   &0.00& 7.75 & 0.03 &  0.26 &  0.62 &  0.08 & 1.03 & 3.01 & 3.82  \\
 \hline
 27.6         &26.38&0.48& 7.96 & 0.02 &  0.20 &  0.58 &  0.19 & 0.93 & 2.86 & 3.51\\
 \hline
28            &27.00 &0.00&8.17 & 0.01 &  0.14 &  0.51 &  0.33 & 0.82 & 2.68 & 3.16\\
\hline
\end{tabular}
%
\end{table*}

For the values of $\mu$ = 26 eV and 27 eV,  the GS is the eigenstate $\ket{\Omega}_{N=26}$, and is a combination 
of $d^7$ ($P_{d^7} \approx 0.3$) and $d^8$ ($P_{d^8} \approx 0.6$).  These state have a non-integer $n_d$ occupation 
due to hybridization of the atomic $d$-states with the substrate. Nevertheless, the ${(\langle N^2 \rangle - {\langle N \rangle}^2)}^{1 \over 2} \approx 0$ 
pointing on the absence of charge fluctuations. The $S$ values lie between of $S=3/2$ (the  atomic $d^7$, $^4F$),  and $S=1$ (the  atomic $d^8$, $^3F$), while the $L$ is close to the atomic $L=3$.
The expectation values of  the $z$-axis projections of the total $\bra{\Omega}J_z\ket{\Omega}$, orbital $\bra{\Omega}L_z\ket{\Omega}$, and spin $\bra{\Omega}S_z\ket{\Omega}$ angular momenta for GS and low-energy  excitation energies for  $\mu =$ 27.0 eV are shown in Tab.~\ref{tab:3}. It is seen that without SOC the GS can be interpreted as $S=1$-like triplet.
For $\mu=28$ eV, the GS is the 
eigenstate $\ket{\Omega}_{N=27}$, and the contributions of $d^7$ ($P_{d^7} \approx 0.1$) and  $d^8$ ($P_{d^8} \approx 0.5$) are reduced while 
$d^9$,  $^2D$ ($P_{d^9} \approx 0.3$) is increased. Again, there are no charge fluctuations near the GS. This GS looks similar to  $S=1/2$ doublet (see Tab.~\ref{tab:3}).

When the SOC is included, for the values of $\mu$ = 26 eV, 27 eV the eigenstate $\ket{\Omega}_{N=26}$  is split to the lowest energy singlet plus excited doublet
(see Tab.~\ref{tab:3}). These states approximately correspond to $\ket{J=1,J_z}$ eigenstates of the effective Hamiltonian~\cite{Tchaplianka2021}, 
\begin{eqnarray}
\hat{H}_{MA} = D \hat{J}_z^2 + E (\hat{J}_x^2 - \hat{J}_y^2 ) \; ,
\label{eq:MA}
\end{eqnarray}
with the uniaxial magnetic anisotropy $D \approx 4.5$ meV, and $E$ = 0. For $\mu=28$ eV, the GS remains  $\ket{\Omega}_{N=27}$ doublet. 

The corresponding densities of $d$-states (DOS)
for the values of $\mu$ = 26 eV, 27 eV, 28 eV are shown in Appendix B Fig.~\ref{fig2}. There is are similarities in the DOS  with and without SOC:
no peak in DOS in a close vicinity of $E_F$. For  these values of $\mu$ and without SOC there are no singlet GS, and no Kondo resonances in the DOS.
In a presence of SOC, even their GS become singlets for  $\mu = 26, 27$ eV, no Kondo peaks are formed. 
For $\mu = 28$ eV the GS solution remains a doublet without Kondo resonance in the DOS.

\begin{table}[!htbp]
\caption{The total number of particles ($d$-shell + bath) $N$, the expectation values $\bra{\Omega}J_z\ket{\Omega}$, $\bra{\Omega}L_z\ket{\Omega}$,$\bra{\Omega}S_z\ket{\Omega}$ angular momenta,
non-zero probabilities $P_{d^n}$ to find the atomic eigenstates $\ket{n}$ with integer occupation $d^{n}$
 for GS and low-energy  excitation energies for different values of $\mu$.}
\label{tab:3}
\centering
\begin{tabular}{cccccccc}
\multicolumn{8}{c}{without SOC}\\
\hline
& Energy (eV) & $J_z$ & $L_z$ & $S_z$ & $P_{d^7}$ & $P_{d^8}$ &$P_{d^9}$\\
\hline
\multicolumn{8}{c}{$\mu$=27.0 eV}\\
$N$=26  & -142.2319 & 0.0 & 0.0 & 0.0  & 0.27 & 0.62 & 0.08 \\
               & -142.2319 & 0.90 & 0.0 & 0.90 & 0.27 & 0.62 & 0.08\\
               & -142.2319 & -0.90 & 0.0 &-0.90 & 0.27 & 0.62 & 0.08\\ 
\multicolumn{8}{c}{$\mu$=27.4 eV}\\
 $N$=26 &-145.3478 & 0.00 & 0.0 & 0.00 & 0.23 & 0.61 & 0.13 \\
              &-145.3478 & 0.81& 0.0 &0.81    & 0.23 & 0.61 & 0.13 \\
              &-145.3478 &-0.81& 0.0 &-0.81   & 0.23 & 0.61 & 0.13 \\ 
 $N$=27 &-145.3490&0.57 &0.0 & 0.57     & 0.19 & 0.55 & 0.23 \\
              &-145.3490 &-0.57 &0.0 & -0.57   & 0.19 & 0.55 & 0.23 \\
\multicolumn{8}{c}{$\mu$=28.0 eV}\\
 $N$=27 &-150.1992 &0.53 &0.0 & 0.53 & 0.14 & 0.51 & 0.33 \\
               & -150.1992 &-0.53 &0.0 & -0.53 & 0.14 & 0.51 & 0.33 \\
               \hline
\multicolumn{8}{c}{with SOC}\\
\hline
& Energy (eV) & $J_z$ & $L_z$ & $S_z$& $P_{d^7}$ & $P_{d^8}$ &$P_{d^9}$\\
\hline
\multicolumn{8}{c}{$\mu$=27.0 eV}\\
$N$=26 &-142.3054 & 0.00 & 0.0 & 0.00  & 0.26 & 0.62 & 0.08\\
              &-142.3009 & 1.48& 0.91 &0.57  & 0.26 & 0.62 & 0.08 \\
              &-142.3009 & -1.48& -0.91 & -0.57 & 0.26 & 0.62 & 0.08 \\ 
\multicolumn{8}{c}{$\mu$=27.6 eV}\\
 $N$=26 &-146.9950 & 0.00 & 0.0 & 0.00 & 0.21 & 0.61 & 0.16 \\
              &-146.9912& 1.10& 0.70 &0.40   & 0.21 & 0.61 & 0.16 \\
              &-146.9912 & -1.10& -0.70 & -0.40 & 0.21 & 0.61 & 0.16 \\ 
 $N$=27 &-146.9931 & 1.43 & 0.95 & 0.48  & 0.18 & 0.54 & 0.26\\
              &-146.9931 & -1.43& -0.95 &-0.48 & 0.18 & 0.54 & 0.26 \\
\multicolumn{8}{c}{$\mu$=28.0 eV}\\
$N$=27 &-150.2373 & 1.37 & 0.91 & 0.45 & 0.14 & 0.51 & 0.33 \\
              &-150.2373 & -1.37& -0.91 & -0.45 & 0.14 & 0.51 & 0.33\\
\hline
 \end{tabular}
\end{table}


\begin{figure}[!htbp]
\centerline{\includegraphics[angle=0,width=1.0\columnwidth,clip]{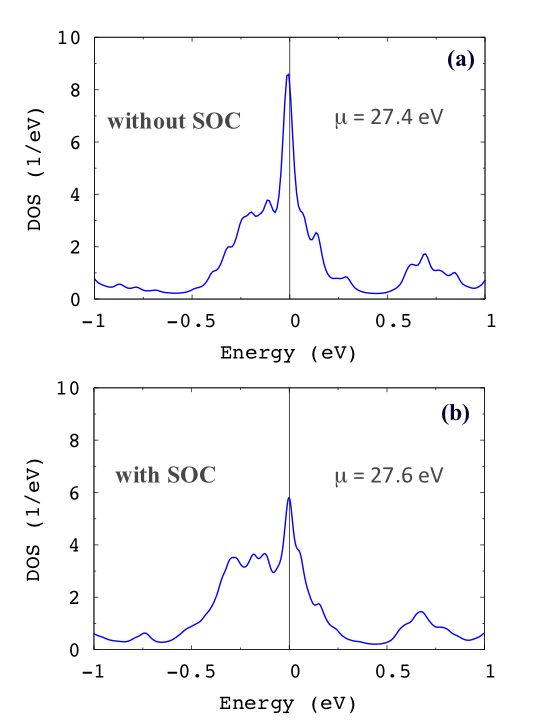}}
\caption{DOS for the Co@Cu(001) as a function of $\mu$= 27.5 eV 
without SOC (a), and with SOC (b),  $\mu$= 27.4 eV without SOC (c), and  27.6 eV with SOC (d).}
 \label{fig3}
\end{figure}

Since the change in the GS with the variation of $\mu$ between 27 eV and 28 eV is observed,
we further adjust the values of $\mu$  in order to keep the same $n_d \approx 8$ without and with the SOC. 
In case of $\mu$=27.4 eV and without the SOC,  we obtain a non-integer $\langle N \rangle$=26.55, non-zero ${(\langle N^2 \rangle - {\langle N \rangle}^2)}^{1 \over 2} \approx 0.5$
charge fluctuations, and $n_d$=7.93. This solution is formally close to  ``$d^8$" state but actually a combination of 
of $d^7$ ($P_{d^7} \approx 0.21$), $d^8$ ($P_{d^8} \approx 0.58$), and $d^9$ ($P_{d^9} \approx 0.18$) atomic states (see Tab.~\ref{tab:2}). 

There is a peak near $E_F$ in the DOS shown in  Fig.~\ref{fig3}(a).
Note that similar peak in DOS was obtained in CTQMC calculations~\cite{Valli2020} without SOC
with the same choice of the Coulomb-$U$ and the exchange-$J$, and
$n_d$=8 very close to our calculations. In Ref.~\cite{Valli2020} it is interpreted as a spectral signature of the Kondo effect.
As follows from  Eq.(~\ref{eq:sdos}) the presence of such a peak signals  the (quasi)-degeneracy of the eigenvalues $E_{N}$,
and $E_{N \pm 1}$. These are the $\ket{\Omega}_{N=27}$ doublet and  $\ket{\Omega}_{N=26}$ triplet states which differ in the energy by 1.2 meV (see Tab.~\ref{tab:3}),
with the doublet GS $\ket{\Omega}_{N=27}$. Since there is no singlet GS, the DOS peak at $E_F$ is not a Kondo resonance, and signals the presence  of valence fluctuations~\cite{Tchaplianka2021}.

When the SOC is included, and with $\mu$=27.6 eV,  there is a non-integer $\langle N \rangle$=26.38, with non-zero 
charge fluctuations ${(\langle N^2 \rangle - {\langle N \rangle}^2)}^{1 \over 2} \approx 0.5$, and $n_d$=7.96 
(see Tab.~\ref{tab:2}). Again, the DOS has a peak at $E_F$ which is shown in Fig.~\ref{fig3}(d). In this case, the  the (quasi)-degeneracy occurs 
between the singlet $\ket{\Omega}_{N=26}$ state being 1.9 meV lower in the energy than  the $\ket{\Omega}_{N=27}$ doublet (see Tab.~\ref{tab:3}).
The  DOS peak at $E_F$ due to $\ket{\Omega}_{N=26}$-to-$\ket{\Omega}_{N=27}$ transition can be interpreted as a Kondo resonance.

For the singlet GS we can use the renormalized perturbation theory~\cite{Hewson} in order to esimate  the Kondo
temperature,
\begin{eqnarray} 
T_K = -{\pi \over 4} Z \Im[\Delta(E_F)] \; ,
\label{eq:TK}
\end{eqnarray}
where 
$$\hat{Z} \approx {Tr[(\hat{I} - d \Re[\Sigma(\epsilon)])/d \epsilon(E_F)]^{-1} A(E_F)] \over Tr[A(E_F)]} $$ 
is a quasiparticle weight, and $A(E_F)$ is the  DOS matrix from Eq.(\ref{eq:sdos}). 
We obtain $Z$=0.097, and corresponding $T_K = 0.019$ eV ($\approx$ 220 K). It exceeds the experimental
estimate $T_K = 88$ K~\cite{Knorr2002} of the Kondo scale. Indeed, Eq.~(\ref{eq:TK}) serves as an order of magnitude
estimate of $T_K$.

The scanning tunnelling spectroscopy measures the differential  conductance $\mathcal{G}(V)$ through the adatom, and 
allows to probe the DOS. Comparison between the experimental and theoretical $\mathcal{G}(V)$ is the most direct way to distinguish between different theoretical
approximations and to identify the most appropriate theoretical approach.
Experimentally $\mathcal{G}(V)$ of Co@Cu(100) was studied in Ref.~\cite{Knorr2002}.
Observed step-like behaviour was interpreted in terms of interference between two tunnelling channels:
(i) tunnelling to the $d$-DOS shown in Fig.~\ref{fig2}, and (ii) tunnelling into the conduction electrons of the
Cu substrate modified by the presence of the Co adatom. At the low bias, the differential conductance is then expressed~\cite{Patton2007}
in the basis of cubic harmonics as, 
\begin{equation}
\mathcal{G}(\omega) \sim \sum_{m} \left( 1 + \Gamma_m ( (1- {q_m}^2) \Im[{G}_{m}(\omega)] + 2 {q_m} \Re [{G}_{m}(\omega)] \right)) \, ,
 \label{eq:didv}
\end{equation}
where ${G}_{m} (\equiv {G}_{mm})$ is a Green's function of the Hamiltonian Eq.(~\ref{eq:hamilt}), $ \Gamma_m \equiv -\Im[\Delta_{m}(E_F)]$ is a hybridization between the $d$-level $m$ and the substrate shown in Fig.~\ref{fig1}C,
and  $q_m$ is a Fano parameter. For the strongly localized Co adatom $d$-orbitals~\cite{Wehling2010}, 
$$q_m \approx - \Re[{G}_{0,m}(E_F)]/ \Im[{G}_{0,m}(E_F)] \; .$$
The calculated $\mathcal{G}(V)$ is in a fair quantitative agreement with the experimental data~\cite{Knorr2002}. 
Note that our results seem to agree with the experiments better than those  of  Ref.~\cite{Lounis2020}.
Contrary to proposal of the Ref.~\cite{Lounis2020}, attempting to explain the zero-bias anomaly in Co@Cu(100) as the results
of inelastic spin excitations, our theory demonstrates that they can be better explained from the point of view of the "Kondo" physics.

\begin{figure}[!htbp]
\centerline{\includegraphics[angle=0,width=1.0\columnwidth,clip]{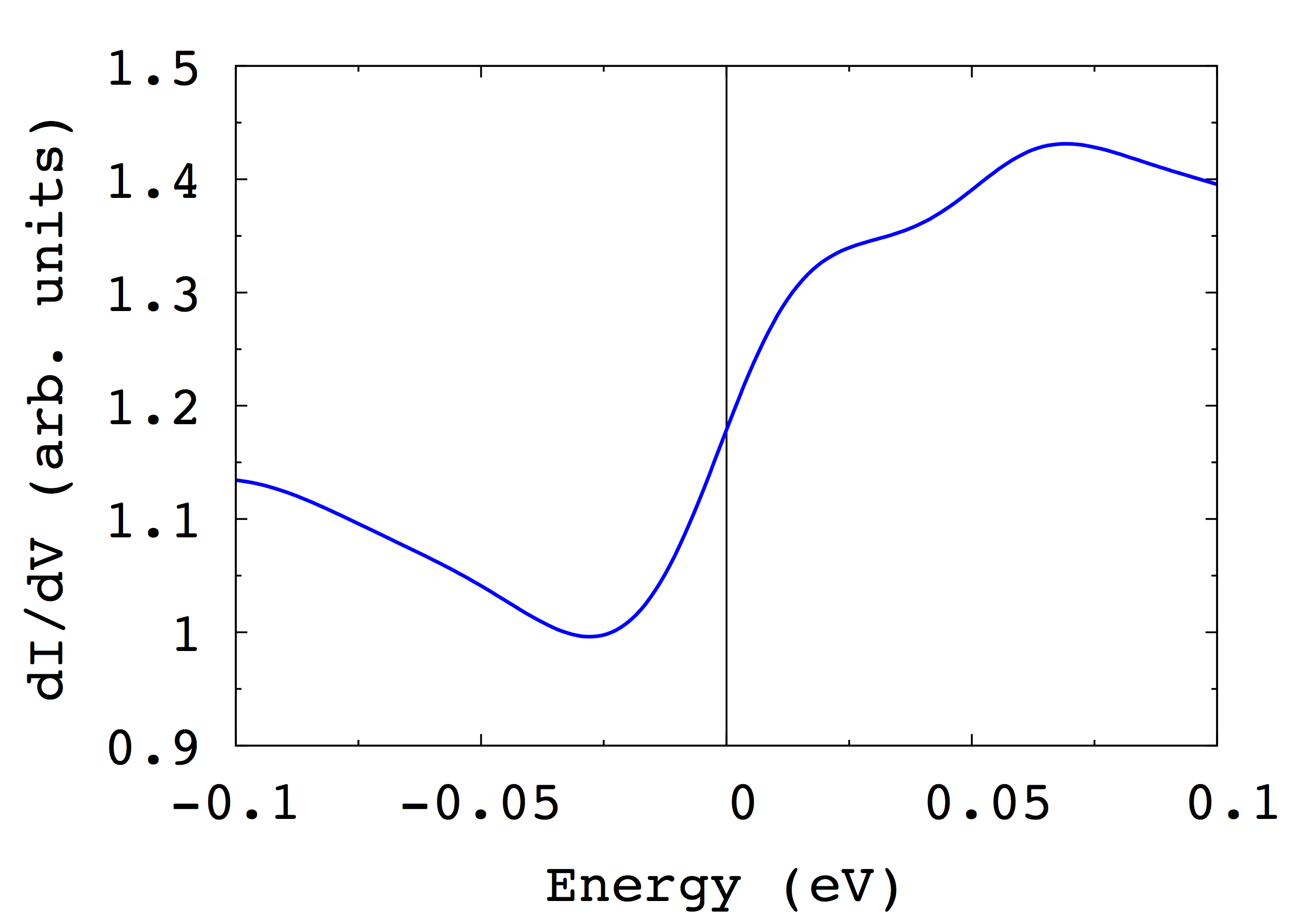}}
\caption{Differential conductance $\mathcal{G}$ calculated making use of the Eq.(~\ref{eq:didv}).}
 \label{fig4}
\end{figure}

\section{Summary} 
The many-body calculations within the multi-orbital SIAM for the Co adatom on the Cu(100) surface are performed.
DFT calculations were used to define the input for the discrete bath model of forty bath orbitals, and the SOC included.
We found that the peak in the DOS at $E_F$ can occur for the Co atom $d$-shell occupation $n_d \approx$ 8, and is connected to 
quasi-degenerate ground state of the SIAM. Without SOC, the lowest energy state is an effective $S=1/2$-like doublet, and next to
it there is an effective $S=1$-like triplet, so the resonance in the DOS($E_F$) does not represent a Kondo resonance.
When SOC is included, the triplet states are split like  $\ket{J=1,J_z}$  eigenstates in a presence of the magnetic anisotropy 
 $\hat{H}_{MA} = D \hat{J}_z^2$, so that the  $\ket{J=1,J_z=0}$  singlet becomes a ground state. The corresponding  DOS($E_F$) peak
corresponds to   the Kondo  resonance. This solution is verified by comparison with experimentally observed   
zero-bias anomaly in the differential conductance.  Our calculations illustrate the essential role which the SOC, and corresponding
uniaxial magnetic anisotropy, is playing in a formation of
Kondo singlet in the multi-orbital low-dimensional systems.  

\section{Acknowledgments} 
Financial support was provided by Operational Programme Research, Development and Education financed by European Structural and Investment Funds and the Czech Ministry of Education, Youth and Sports 
(Project No. SOLID21 - CZ.02.1.01/0.0/0.0/16$_{-}$019/0000760),
and by the Czech Science Foundation (GACR) grant No. 21-09766S. 
The work of A.I.L. is supported by European Research Council via Synergy Grant 854843 - FASTCORR.

\clearpage

\appendix
\section{Fitting the bath hybridization}
With the specific choice of the Cartesian reference frame (see Fig. 1), the  local Green's function $G_0(z)$ becomes diagonal in 
the basis of cubic harmonics $m = \{xz, yz, xy, x^{2}-y^{2}, 3z^{2}-r^{2}\}$.  Moreover, it is convenient to use the imaginary energy axis over the Matsubara frequencies $i\omega_n$.
The corresponding non-interacting Green's function of the Eq.(\ref{eq:hamilt}) will then become 
$$
{G}_{0,m}\left(i \omega_n\right) = \frac{1}{i\omega_n-\epsilon_{m} -\Delta_{m}\left(i\omega\right)},
$$
with the hybridization function

\begin{eqnarray}
\Delta_{m}\left(i\omega_{n}\right) & =i\omega_{n}-\epsilon_{m} - {G}_{0,m}^{-1}\left(i\omega_{n}\right).
\label{eq:D}
\end{eqnarray}

Thus, the hybridization function Eq.~(\ref{eq:D}) can be evaluated making use of the local Green's function $G_0(z)$.
The discrete bath model is built by finding bath energies and
amplitudes which reproduce the continuous hybridization function as
closely as possible.

\begin{equation}
\tilde{\Delta}_{m}\left(i\omega_n\right)  =\sum_{k=1}^{K}\frac{V_{km}^{2}}{i\omega_n-\epsilon_{km}}.
\end{equation}

The fitting is done by minimizing the residual function,  
\begin{eqnarray}
f_m\left(\left\{ \epsilon_{km},V_{km}\right\} \right) & =\sum_{n=1}^{N_{\omega}}\frac{1}{\omega_{n}^{\gamma}}\left|\tilde{\Delta}_m\left(i\omega_{n}\right)-\Delta_m\left(i\omega_{n}\right)\right|^{2}, \nonumber \\
\end{eqnarray}
using the limited-memory, bounded Broyden--Fletcher--Goldfarb--Shanno
method~\cite{zhu1997,Morales2011}, with the parameters $\epsilon_{km}$ and $V_{km}$ as variables. The factor ${1} \over {\omega_{n}^{\gamma}}$ with $\gamma=0.5$ is used to attenuate the
significance of the higher frequencies. 

\begin{table}[!htbp]
\centering
\caption{Values of the $d$ shell $\Delta_{\rm CF}$ (eV), the bath energies $\epsilon^{k}_{m}$ (eV), and hybridisation parameters $V^{k}_{m}$ (eV) 
evaluted from LDA .}
\label{tab:1} 
\begin{tabular}{cccccc}
$m$& \multicolumn{1}{c}{$xz$} & \multicolumn{1}{c}{$yz$}&\multicolumn{1}{c}{$xy$}&  \multicolumn{1}{c}{$x^2-y^2$}&\multicolumn{1}{c}{$3z^2 - r^2$} \\ \hline
$\Delta_{CF}$    &-0.043   &-0.043      &0.117 & 0.053 &-0.082 \\ \hline
$\epsilon_{k=1,m}$ & -2.16& -2.16 & -1.99 & -2.01 & -2.57 \\ 
$V_{k=1,m}$ & 0.85 & 0.85 & 0.65 & 0.65&0.72 \\ 
\hline
$\epsilon_{k=2,m}$ & -0.08 & -0.08& 0.001 & -0.02 & -0.05 \\ 
$V_{k=2,m}$ & 0.18 & 0.18 & 0.08 & 0.10 &0.13  \\ 
\hline
$\epsilon_{k=3,m}$ & 0.51 &  0.51 & 1.45 & 0.53 & 0.43 \\ 
$V_{k=3,m}$ & 0.36 & 0.36 & 0.55 & 0.34 &0.32 \\ 
\hline
$\epsilon_{k=4,m}$ & 7.56 & 7.56 & 7.80 & 8.16 & 7.72 \\ 
$V_{k=4,m}$ & 2.08 & 2.08 & 2.12 & 1.78 &1.70  \\ 
\hline
\end{tabular}
\end{table}
\pagebreak

\section{DOS as a function of $\mu$ for Co on Cu(001)}

\begin{figure}[!htbp]
\centerline{\includegraphics[angle=0,width=1.0\columnwidth,clip]{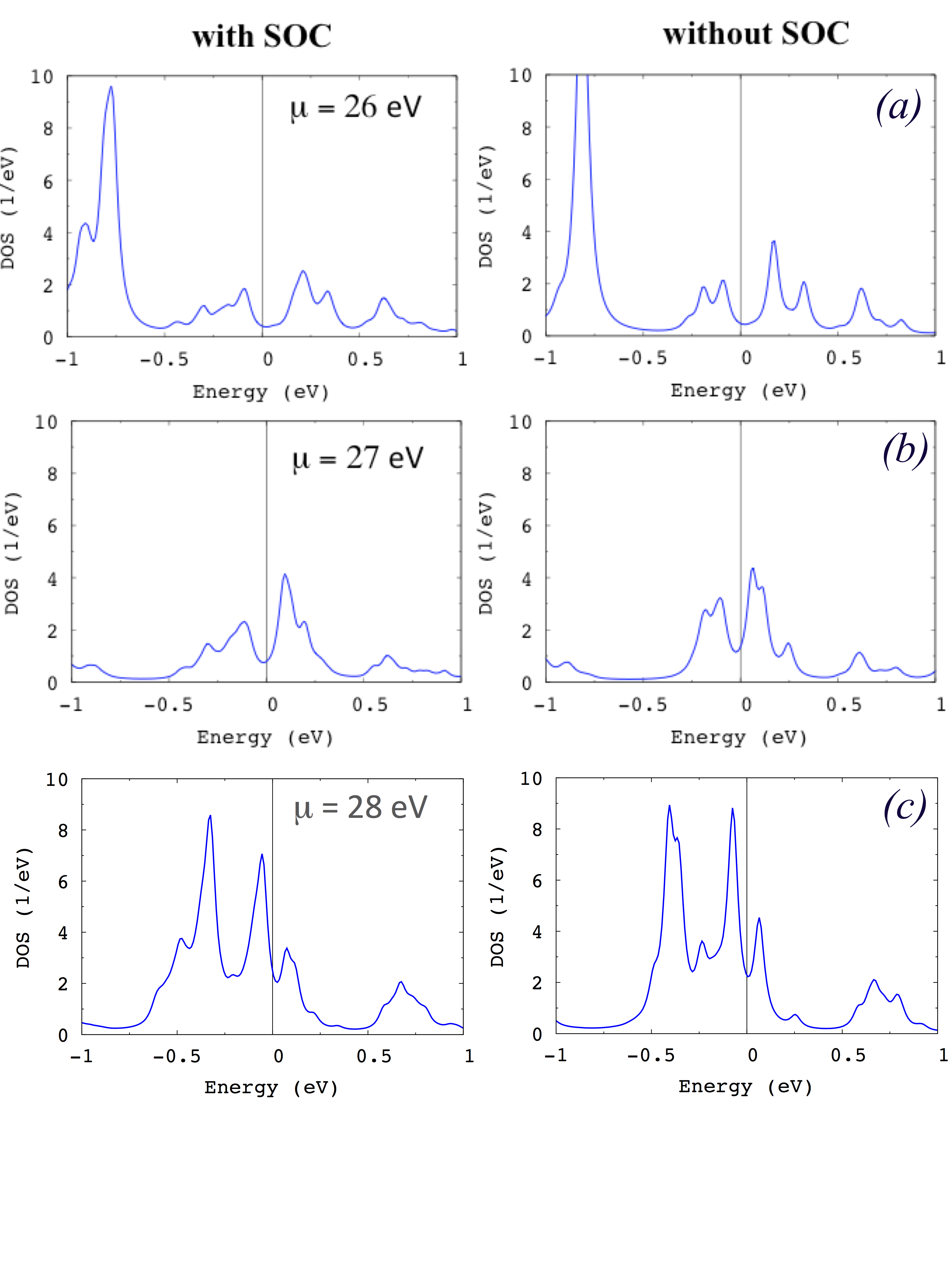}}
\caption{DOS for the Co@Cu(001) with and without SOC as a function of $\mu$= 26 eV (a), 27 eV (b), and 28 eV (c)}
 \label{fig2}
\end{figure}

\clearpage
\bibliographystyle{iopart-num}

\end{document}